\documentclass{elsart}

 \usepackage{graphicx}
 \usepackage{epsfig}

\usepackage{amssymb}

\usepackage[english,francais]{babel}

\newtheorem{e-proposition}[theorem]{Proposition}

\newtheorem{e-definition}[theorem]{Definition\rm}

\def\og{\leavevmode\raise.3ex\hbox{$\scriptscriptstyle\langle\!\langle$~}}
\def\fg{\leavevmode\raise.3ex\hbox{~$\!\scriptscriptstyle\,\rangle\!\rangle$}}

\begin{document}

\begin{frontmatter}


\selectlanguage{english}
\title{Can cooperation slow down emergency evacuations?}


\selectlanguage{english}
\author[authorlabel1]{Emilio N. M. Cirillo},
\ead{emilio.cirillo@uniroma1.it}
\author[authorlabel2]{Adrian Muntean}
\ead{a.muntean@tue.nl}

\address[authorlabel1]{Dipartimento di Scienze di Base e Applicate per 
             l'Ingegneria, Sapienza Universit\`a di Roma, 
             via A.\ Scarpa 16, I--00161}
\address[authorlabel2]{Department of Mathematics and Computer Science, CASA -- Center for Scientific Computing and Applications, ICMS -- Institute for Complex Molecular Systems, Eindhoven University of Technology, PO Box 513, 5600 MB, Eindhoven, The Netherlands}


\medskip
\begin{abstract}
We study the motion of pedestrians through obscure corridors where the lack of visibility hides the precise position of the exits. 
Using a lattice model, we explore the effects of 
cooperation  on the overall exit flux (evacuation rate). More precisely, we study the effect of the buddying threshold 
(of no--exclusion per site) on the dynamics of the crowd.  In some cases, we note that  if the evacuees tend to cooperate and act altruistically, then their collective action tends to  favor the occurrence of disasters. 

\vskip 0.5\baselineskip

\selectlanguage{francais}
\noindent{\bf R\'esum\'e}
\vskip 0.5\baselineskip
\noindent
Nous \'etudions la dynamique des mouvements de foules dans des corridors dont la visibilit\'e est tres r\'eduite. Tout en particulier, nous nous int\'eressons \'a des corridors dont les sorties ne sont pas visibles. \'A l'aide de notre mod\`ele -- un automate cellullaire -- nous exploitons les effets que la cooperation parmi les pi\'etons  produit sur le flux macroscopique d'\'evacuation. Dans des certains cas, nous observons que  si les pi\'etons se comportent altruistiquement, alors  des ph\'enom\`enes macroscopiques catastrophiques \'emergent de la combinaison de ces interactions locales.  


\vskip 0.5\baselineskip
\noindent{\small{\it Mots-cl\'es:} La dynamique des mouvements de foules; des automates cellulaire; un sc\'enario d'\'evacuation}}
\end{abstract}
\end{frontmatter}


\selectlanguage{english}
\section{Introduction}
This Note studies the following evacuation scenario: A large group of people needs to evacuate a subway station or a tunnel system [with complicated geometry] without visibility. The lack of visibility, or say, the heavily reduced visibility, can be imagined to be due to the breakdown of the electricity network, or due to the presence of a very dense or irritating smoke. We assume also that the evacuation audio signaling is not activated and that, in spite of all these difficulties,  all pedestrians need to travel through this dark region and must  find as soon as possible their way out towards the hidden exit. Additionally,  we assume that all the persons are {\em equally fit} (i.e. they are indistinguishable) and that none of them  has {\em a priori} knowledge on the location of the exit. To keep things simple, we consider that there are not spatial heterogeneities inside the region in question.


There are studies done [especially for fire evacuation scenarios] on how information and way finding systems are perceived by individuals.  One of the main questions in fire safety research is whether green flashing lights can influence the evacuation (particularly, the exit choice); see e.g. \cite{Nilsson,Shields,Bauke} (and the fire engineering references cited therein) and \cite{Yuan} (partial visibility due to a 
non--uniform smoke concentration) \cite{Jun} (partial visibility as a function of smoke's temperature), \cite{Zheng} (flow heterogeneity due to fire spreading). If exits are visible, then an impressive amount of literature provide proper working methodologies and efficient simulation tools. 
  Preliminary assessment tests (cf. \cite{PASS,Bryan,Bruno}, e.g,) and many modeling approaches  (deterministic or stochastic) succeed to capture qualitatively basic behaviors of humans (here referred to as pedestrians ) walking within a given geometry towards  {\em a priori} prescribed exits; see, for instance,  social force/social velocity crowd dynamics models (cf. e.g.  \cite{HelbingMolnar}, \cite{PiccoliTosinMeasTh}, \cite{EversMuntean}, \cite{Degond}), simple asymmetric exclusion models (see chapters 3 and 4 from \cite{Schadschneider2011} as well as references cited therein), cellular automaton-type models \cite{Kirchner,Guo}, etc.

But, as far as we are aware,  nothing seem to be known on evacuating people through regions without visibility, therefore our interest. 

By means of a minimal model, we wish to describe how a bunch of people located inside a dark (smoky, foggy, etc.) corridor  exits through 
an invisible door open in one of the four walls. We decide on this way (cf. section \ref{s:modello}) on a possible mechanism regarding how do pedestrians choose their path and speed when they are about  to move through regions with no visibility. The question that triggers our attention here is the following:
\begin{center} 
 {\em Is cooperation/group formation the right strategy to choose  to ensure the crowd evacuation 
within a reasonable time?}
\end{center}

\section{A lattice model}
\label{s:modello}
We use a minimal lattice model, which we name the 
{\em reverse mosca cieca game},
where we incorporate a few basic rules for the pedestrians motion in dark.

\par\noindent
\subsection{Basic assumptions on the pedestrians motion}
\par\noindent
We take into consideration the following four mechanisms:
\begin{itemize}
\item[(A1)]
In the core of the corridor, 
people move freely without constraints;
\item[(A2)]
The boundary is reflecting, possibly attracting;
\item[(A3)]
People are attracted by bunches of other people up 
to a threshold, say $T$;
\item[(A4)] 
People are blind in the sense that there is no drift (desired velocity) leading them 
towards the exit.
\end{itemize}

(A1)--(A4) intend to describe the following situation: 

Since, in this framework, neighbors (both individuals or groups) can not be visually identified by the individuals in motion, basic mechanisms like attraction to a group, tendency to align, or social repulsion are negligible and individuals have to live with ``preferences". 
Essentially, their motion is more {\em behavioral} than {\em rational}. We assume that the individuals move freely inside the corridor but they like to buddy to people they accidentally meet at a certain point (site). The more people are localized at a certain site, the stronger the preference to attach to it.  However if the number of people at a site reaches a threshold, then such site becomes 
not attracting 
for eventually new incomers. (A3), referred here as the {\em buddying mechanism}, is the central aspect of our research. 

Once an individual touches a wall, he/she simply felts the need to stick to it at least for a while, i.e. until he/she  can attach to an interesting site (having conveniently many hosts) or to a group of unevenly occupied sites or the exact location of the door is detected. 

Since people have no desired velocity, their diffusion (random walk) together with the buddying are the only transport mechanisms. Can these  eventually lead to evacuation? How efficient is such combination?

In the following, we study the effect of the threshold 
(of no--exclusion per site) 
on the overall dynamics of the crowd.  Here we  describe our results in terms of the averaged outgoing flux; see Fig. \ref{f:uno} and Fig. \ref{f:due}. In a forthcoming publication, we will investigate also  other macroscopic quantities like the stationary occupation numbers  and  stationary correlations.
 
\begin{figure}[htbp]
\begin{center}
\begin{picture}(500,200)(-40,350)
\resizebox{16cm}{!}{\rotatebox{0}{\includegraphics{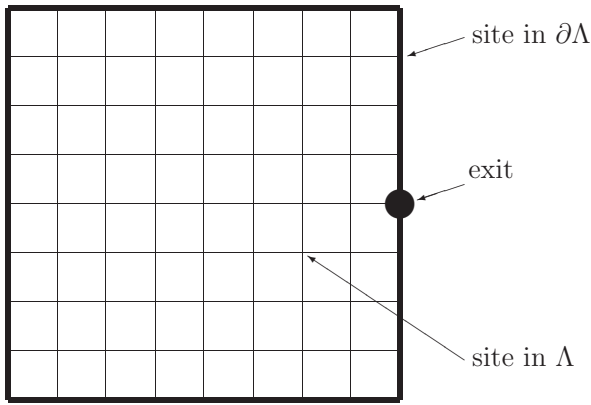}}}
\end{picture}
\caption{Geometry of the model.}
\label{f:zero}
\end{center}
\end{figure}

\subsection{The lattice model}
We start off with the construction of the lattice.  
Let $\Lambda\subset\mathbb{Z}^2$ be a finite square with 
odd side length $L$. We refer to this generic room as  the {\em corridor} where the dynamics is about to happen. 
Each element $x$ of $\Lambda$ will be called a {\em cell} or {\em site}.
Two sites are said to be {\em nearest neighbor} if and only if their 
mutual Euclidean distance is equal to one. 
For any $x\in\Lambda$, we let $\Gamma(x)$ be the cross--shaped 
subset of $\Lambda$ made of $x$ and its four nearest neighbors. 
The {\em external boundary} $\partial\Lambda$ of $\Lambda$, 
i.e., the collections 
of the sites in $\mathbb{Z}^2\setminus\Lambda$ neighboring one site 
in $\Lambda$, is made of four segments 
made of $L$ cells each. 
The point at the center of one of these four sides is called {\em exit} and is 
denoted by $x_\textrm{e}$.

Let $N$ be a positive integer denoting the (total) 
{\em number of individuals} inside 
the corridor $\Lambda$. 
We consider the state space $X:=\{0,\dots,N\}^\Lambda$.  
For any state $n\in X$, we let 
$n(x)$ be the {\em number of 
individuals} at cell $x$. 

We define, now, a Markov chain $n_t$ on the finite state space 
$X$ with discrete time $t=0,1,\dots$. For any $x\in\Lambda$, 
$n_t(x)\in\{0,1,\dots,N\}$ is the number of individuals at site 
$x$ and at time $t$. 
At each time $t$, the position of all the individuals 
on each cell is updated according to the following rule: 
the individual at site $x\in\Lambda$ jumps to the site 
$y\in\Lambda\cup\{x_\textrm{e}\}$ with the probability $p(x,y)$ that 
will be defined below; note it can be $y=x$. 
If one of the individuals jumps on the exit 
cell a new individual is put on a cell of $\Lambda$ 
chosen randomly with the uniform probability $1/L^2$. 

The dynamics is controlled by a single integer parameter $T$, 
called {\em buddying threshold}, which has to be chosen 
in the set $\{0,1,\dots,N\}$. 
We define the function $S:\mathbb{N}\to\mathbb{N}$ 
such that for any $k\in\mathbb{N}$
\begin{displaymath}
S(k)=1
\textrm{ if } k>T
\;\;\textrm{ and }\;\;
S(k)=k+1 
\textrm{ if } k\le T.
\end{displaymath}
Note that, whatever the value of $T$ is,
for $k=0$ we have $S(0)=1$.

Given a configuration $n\in X$, 
given $x\in\Lambda$ and $y\in\Lambda\cup\{x_\textrm{e}\}$, 
we define 
the probability $p(x,y)$ for an individual to jump 
from $x$ to $y$ as follows: 
We let $p(x,y)=0$ if $y\not\in\Gamma(x)$ and 
\begin{displaymath}
p(x,y):=
\frac{{\displaystyle S(n(y))}}
     {{\displaystyle \sum_{w\in\Gamma(x)\cap\Lambda}S(n(w))}}
\end{displaymath}
for any $y\in\Gamma(x)$, 
where we understand $n(x_\textrm{e})=T+1$.
With this choice we treat the exit as a bulk site 
at the threshold. 
It is worth stressing here that $T$ is not a threshold in 
$n(x)$ -- the number of individuals per cell. 
It is a threshold in the probability that 
such a cell is likely to be occupied or not.  
 
Note that the 
approach we take here is very much influenced by a basic scenario described 
in \cite{Emilio02,Emilio} for randomly moving sodium ions willing to pass through a switching 
on--off  membrane gate. The major difference here is twofold: the gate is permanently open and the buddying principle is activated.

\section{Comments on cooperation effects -- the buddying threshold $T$}
\label{s:commenti}
%
%
%

The possible choices for the parameter $T$ correspond to two different physical 
situations. The first one, for $T=0$,  the function $S(k)$ is equal to $1$ 
(the minimal quantum) 
whatever the occupation numbers 
are. This means that
each individual has the same probability to jump to one of its nearest neighbors
or to stay on his site. This is resembling  the independent
symmetric random walk case; the only difference is that with the same probability 
the individuals can decide not to move. We expect that this ``rest probability" just changes a little 
bit the time scale. 

The second physical case is $T>0$. For instance,  $T=1$ means {\em mild buddying}, while 
$T=30,100$ would express an {\em extreme buddying}. 
Note that $T=100$ is not realist but we used it in the simulations 
to check the behavior of the model with increasing $T$. 
No simple exclusion is included in this model:
on each site one can cluster as many particles (pedestrians) as one wants. 
The basic role of the threshold is the following: 
The weight associated to the jump towards the site $x$ 
increases from $1$ to $1+T$ proportionally to the occupation 
number $n(x)$ until $n(x)=T$, after that level it drops back to $1$. 
Note that this rule is given on weights and not on probabilities. Therefore,  
if one has $T$ particles at $y$ and $T$ at each of its nearest 
neighbors, then  at the very end one will have that the probability to stay 
or to jump to any of the nearest neighbors is the same. Differences 
in probability are seen only if one of the five (sitting in the core) sites 
involved in the jump (or some of them) has an occupation number 
large (but smaller than the threshold). 
 
\begin{figure}[htbp]
\begin{center}
\begin{picture}(500,200)
\put(-60,160)
{
\resizebox{8cm}{!}{\rotatebox{-90}{\includegraphics{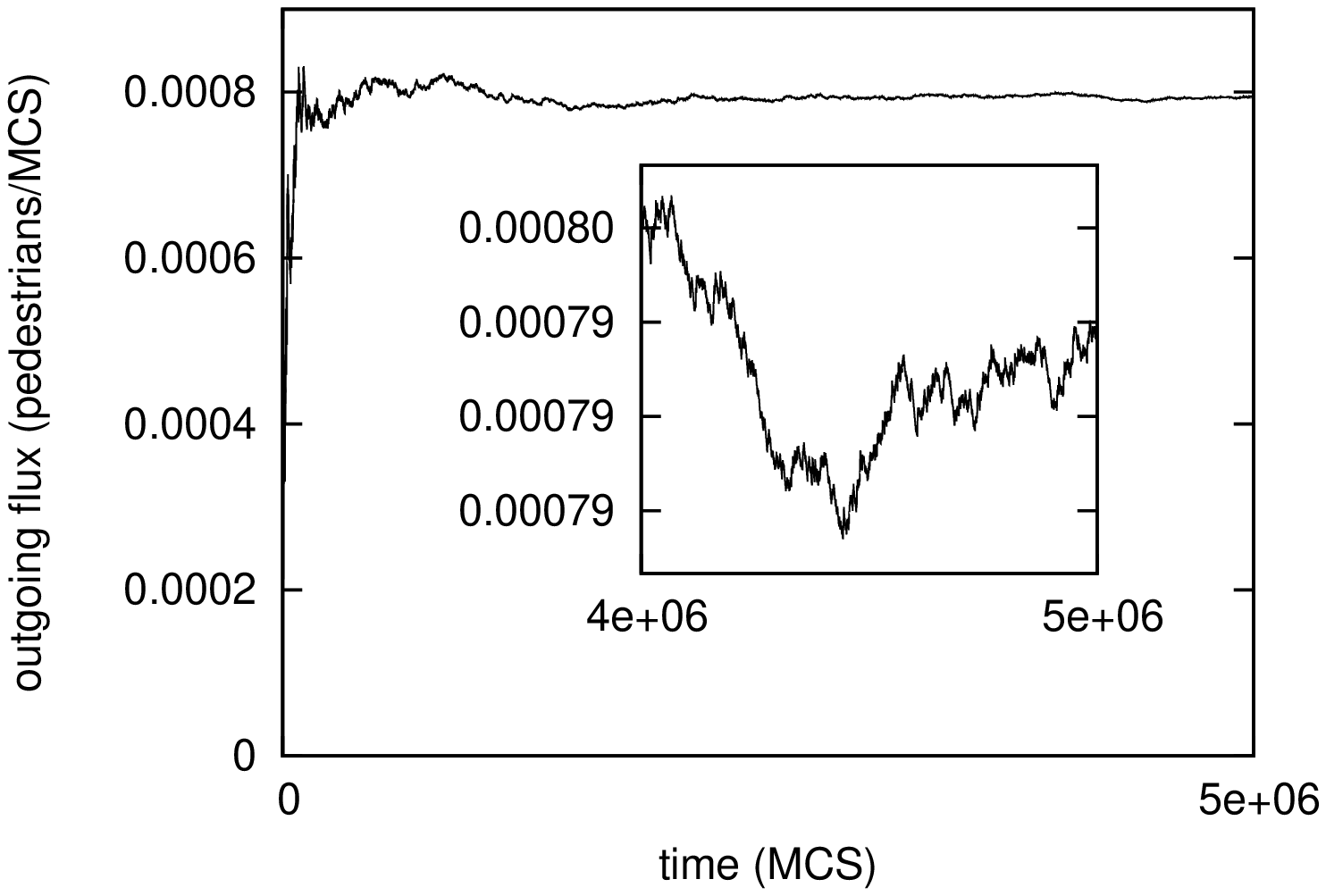}}}
}
\put(180,160)
{
\resizebox{8cm}{!}{\rotatebox{-90}{\includegraphics{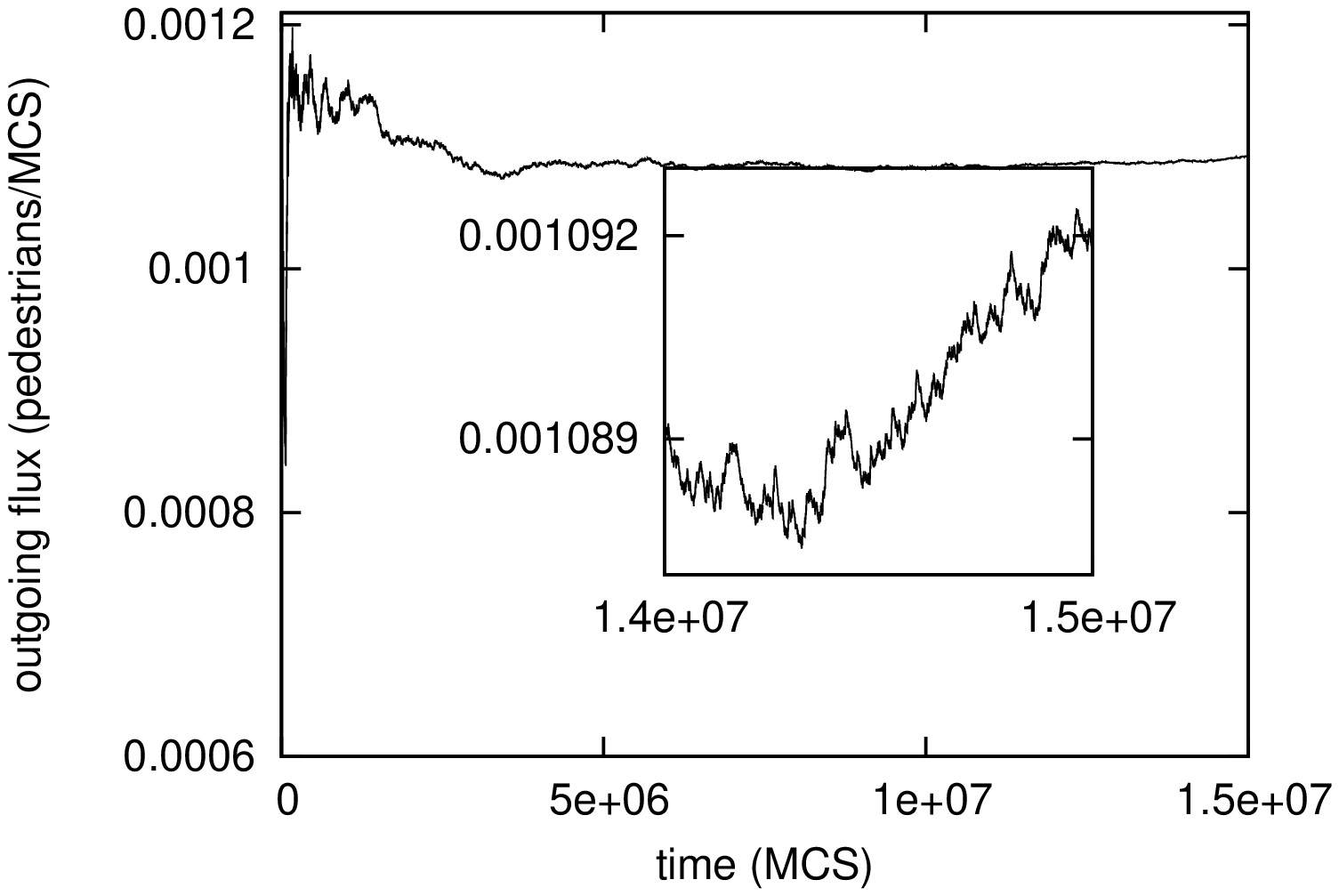}}}
}
\end{picture}
\caption{Averaged outgoing flux vs. time in the case 
$T=0$ and $N=100$ on the left and 
$T=100$ and $N=100$ on the right. 
The inset is a zoom in the time interval 
$[4\times10^6,5\times10^6]$ on the left and 
$[1.4\times10^7,1.5\times10^7]$ on the right.}
\label{f:uno}
\end{center}
\end{figure}

\begin{figure}[htbp]
\begin{center}
\includegraphics[angle=-90,width=0.75\textwidth]{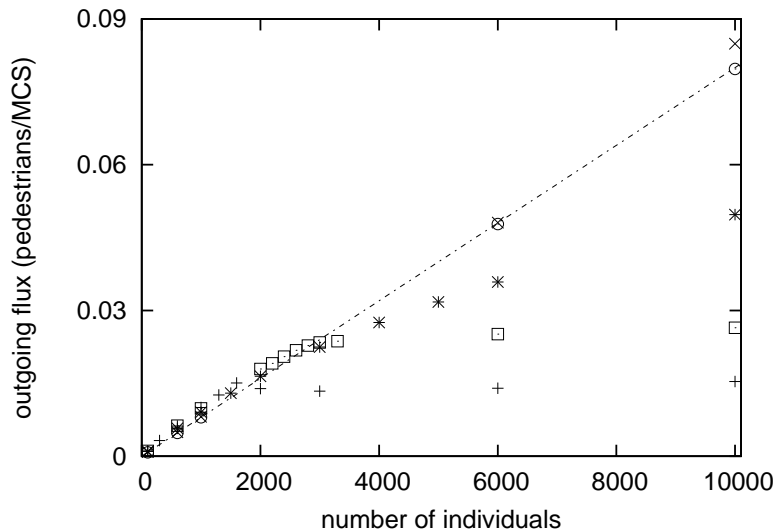}\\
\caption{Averaged outgoing flux vs. number of pedestrians.  
The symbols $\circ$, $\times$, $*$, $\square$, and $+$ 
refer respectively to the cases $T=0,1,5,30,100$.
The straight line has slope $8\times10^{-6}$ and has 
been obtained by fitting the Monte Carlo 
data corresponding to the case $T=0$.}
\label{f:due}
\end{center}
\end{figure}

The main quantity of interest is the {\em outgoing flux}, namely, the 
total number of pedestrians which reached the exit in a time interval 
 times the length of the time interval (number of 
Monte Carlo Steps (MCS)). 
In Fig.~\ref{f:uno} we plot the outgoing flux as function of time: 
note that the measured quantities approach a reasonably stationary 
value after about $10^7$ Monte Carlo steps. 
In Fig.~\ref{f:due}, we see that the overall dynamics very 
much depends on both the number $N$ of individuals and their 
ability to cooperate (the threshold $T$). In particular, this 
figure indicates that if $N$ is sufficiently large, then 
cooperation does not seem to be the best option. Otherwise, for $N$ 
sufficiently small, cooperation seems to be able to ensure a timely 
evacuation. This counter intuitive effect is not explaining why 
cooperation can, under certain circumstances, slow down emergency evacuations. More research initiatives in this direction are needed in order to shed light on this important issue.

\section*{Acknowledgments} We thank J.-J. Oosterwijk, E. Tosolini, M. B\"ohm, E. Ronchi, and G. Th\'eralauz for fruitful discussions on this and closely related topics.
\bibliographystyle{plain}
\bibliography{smoke}

\end{document}